\documentclass[5p]{elsarticle}
\usepackage[frozencache]{minted}
\usepackage[english]{babel}
\usepackage{amsmath}
\usepackage{graphicx}
\usepackage{url}

\journal{Astronomy and Computing, Elsevier}

\begin{document}

 \begin{frontmatter}

\title{Kliko - The Scientific Compute Container Format}

\author[ska,rhodes]{Gijs Molenaar}
\author[ska,rhodes]{Spheshile Makhathini}
\author[julien,ska]{Julien N. Girard}
\author[rhodes,ska]{Oleg Smirnov}

\address[ska]{SKA SA, Cape Town, South Africa}
\address[julien]{AIM/CEA-Saclay, Université Paris Diderot, France}
\address[rhodes]{Department of Physics \& Electronics, Rhodes University, Grahamstown, South Africa}

\date{\today}

\begin{abstract}
Kliko is a Docker-based container specification for running one or multiple related compute jobs. The key
concepts of Kliko are the encapsulation of data processing software into a container and the formalization of the input,
output and task parameters. By formalizing the parameters, the software is represented as abstract building blocks
with a uniform and consistent interface. The main advantage is enhanced scriptability and empowering pipeline composition.

Formalization is realized by bundling a container with a Kliko file, which describes the
IO and task parameters. This Kliko container can then be opened and run by a Kliko runner. The Kliko runner will parse
the Kliko definition and gather the values for these parameters, for example by requesting user input or retrieving
pre-defined values from disk. Parameters can be various primitive types, for example: float, int or the path to a file.

This paper will also discuss the implementation of a support library named Kliko which can be used to create Kliko containers,
parse Kliko definitions, chain Kliko containers in workflows using a workflow manager library such as
\textit{Luigi}. The Kliko library can be used inside the container to interact with the Kliko runner.

Finally to illustrate the applicability of the Kliko definition, this paper will discuss two reference implementations
based on the Kliko library: RODRIGUES, a web-based Kliko container scheduler and output visualizer specifically for astronomical
data, and VerMeerKAT, a multi-container workflow data reduction pipeline which is being used as a prototype pipeline for
the commissioning of the MeerKAT radio telescope.

The Kliko library is open source. The documentation and source code can be found on the main website.\footnotemark
\end{abstract}


\end{frontmatter}

\footnotetext{\url{https://github.com/gijzelaerr/kliko}}

\section{Introduction}
\subsection{Software in science}
The use of computer software in research has resulted in significant hardware and software developments
in computing science. Nowadays, the number of different scientific software packages is overwhelming, and it has become
progressively difficult for users (e.g. a scientist) to evaluate the relevance, usage and the performance of these packages.

Firstly, installing scientific software can be cumbersome, especially when the installation
and/or compilation is poorly designed. The software code, the library dependencies, the host platform and the compilers
may change over time, making it unclear how the original developer(s) intended to install and use the software. Secondly,
conflicting dependencies may arise when different software packages are built together, making it difficult to install them on
the same system. Thirdly, software packages have non-uniform interfaces as they have varying expectations of interaction
with a user or with other packages on the same system.

Kliko is a Docker-based encapsulating and chaining framework that purports to mitigate these issues
by creating a container of the software thereby solving the first and second issue above. The third issue can then
be solved by building a Docker container that has minimal extra requirements, i.e. the Kliko definition.

Kliko consists of two parts: i) a set of utilities for creating a container, including parsers to
check if all (meta) data is valid; and ii) a support library that can be used to schedule a Kliko
container and run it from a command line or from a web interface.

Kliko is not a pipeline construction tool itself, nor a web interface, but it can assist in making these.

\subsection{Software containerization with Docker}
Containerization is a method for building self-contained environments (called ``containers'') for applications. These
containers can then be distributed and used with minimal effort on a large variety of platforms.

Containerizing applications is not new. Similar techniques have been applied before, e.g. jail for FreeBSD
\footnote{\url{https://www.freebsd.org/doc/handbook/jails.html}}, zones for Solaris \cite{PriceSolaris} and chroot for
GNU/Linux \footnote{\url{http://man7.org/linux/man-pages/man2/chroot.2.html}}. However, their application was mostly
limited to enhancing security and to carry out clean builds of the UNIX system. The addition of operating system (OS)
level process isolation, named control groups or cgroups\cite{rosen2013resource}), to the popular Linux kernel
(since 3.8, 2008) accelerated the adoption of containerization for the usage of software distribution.

There are multiple software projects leveraging cgroups, for example
rkt\footnote{\url{https://github.com/coreos/rkt}}, Docker\cite{Boettiger14}\footnote{\url{https://www.docker.com}},
Singularity\cite{kurtzer2016}\footnote{\url{http://singularity.lbl.gov}} and
LXC\footnote{\url{https://linuxcontainers.org}}. Docker \cite{Merkel2014} is currently the most popular container
technology with the largest community of users and the most momentum for future development and support. Kliko aims to
be agnostic of the container technology, but since Docker has the biggest user community, we focus on this
implementation.

In Docker, an image is built using a initialization script (a ``Dockerfile'') which contains the recipe
to install or build the application. The Dockerfile is a series of commands applied to a basic and clean Docker image,
typically a headless Linux distribution. These base images are retrieved from an online database provided by Docker,
and stored locally. The Dockerfile, when
executed, will create an ``image'' which is a ``inactive'' snapshot of the virtualized application. An image becomes a
container when instantiated (e.g. the application runs). The difference between active and inactive is important, a
container is an image with an unwritten (dirty) state.

An application that is containerized is self-contained and can be seen as a complete OS without kernel. The container
could even only contain a statically compiled binary, but in practise it is useful to have the tools and package
manager of a Linux distribution available inside the container. Theoretically, to run a Docker container using Docker
on a host machine, the only requirement is to have the Docker daemon running on the host. Unfortunately, there are
some hardware specific edge cases like CPU register usage optimization and GPU acceleration. These cases will be
discussed in section \S\ref{sec:Limitations}.

A Docker container ``image'' is basically a file system snapshot of a minimal OS. The ``target'' application (i.e.
the one to host) and its library dependencies are installed inside this virtual isolated file system. When the
application is started, the container file system is exposed to the application as the working environment.

On a kernel level, cgroups and namespaces are used to create a new isolated environment for the application, limiting
access to other processes on the host and presenting the isolated environment as if is a separate host to the application.
Intuitively, this can be seen as similar technology as CPU level OS virtualization like VirtualBox, but in the case of
containers, the kernel is shared by the host and the guest.

A Docker container also gets a private IP address on an internal network range. This makes the container appear as a
separate networked machine to the host. By default, access to network ports are restricted and access needs to be granted per
port. One can also forward the port to an external interface where it will appear as the service is running on the host
itself.

All of the above might appear similar to simple virtualization, but containerization has some clear additional advantages.
Firstly, when using Docker, available physical resources do not need to be
partitioned between the host and the guest. While memory size allocated for a virtual machine is fixed or not easy to change,
running containers does not require the user to fix this memory size, although it still is possible to limit the amount of memory
allocatable by the process. Secondly, there is no CPU instruction emulation, as the process is directly executed on the
host kernel. Thirdly, there is minimal startup and shutdown overhead for starting containers as the containerized OS is
reduced to minimal consumption. Startup time is instantaneous (in the millisecond range) and loading time will only
become noticeable when high numbers of containers are spawned.

In addition to containerization, Docker also offers other features: it uses an ``union'' file system to join
multiple layers of file systems together. The intermediate result of each command in the Dockerfile is cached and stored
in layers. These layers can be reused by other containers, allowing data sharing between them, which reduces the size of
the storage requirements. These layers can also be stored in a central location, where they can be distributed and reused
in both a public or private way.

\section{The Kliko specification}

The Kliko specification is designed to extend containerization with an uniform interface resulting in simplified
interaction with the containerized application.

The Kliko specification describes how a Kliko container should look like and what a Kliko container should expect during
runtime. The relevant terminology is listed below:

\underline{Def~1:} The Kliko Image\\
A Docker image complying to the Kliko specification. An image is a read-only ordered collection of root file system
changes and the corresponding execution parameters for use within a container runtime.

\underline{Def~2:} The Kliko Container\\
A container in an active (or inactive if exited) stateful instantiation of a Kliko image. 

\underline{Def~3:} The Kliko Runner\\
A process that can run a Kliko image to make a container. For example the Kliko-run command line tool, or RODRIGUES
(see \S \ref{rodrigues}).

\underline{Def~4:} The Kliko Parameters\\
A list of parameters that can influence the behavior of the software in the container. The list can be arbitrary in
size and consists of any combination of primitive types listed in Table~\ref{tab:klikotypes}.

\subsection{The Kliko Image}
A Kliko image should contain a \texttt{/kliko.yml} file in YAML\footnote{\url{http://www.yaml.org}}
syntax following the Kliko schema \ref{klikoformat}. YAML is a human-readable data
serialization language and stands for \textit{YAML Ain't Markup Language}.
The Kliko image should also contain a \texttt{/kliko} file which is called during runtime by the Kliko runner. This
Kliko script can be anything executable, but in most cases, it will be a Python script using the Kliko library to check
and parse all related Kliko tasks during runtime. Note that we have deliberately chosen not to use the ENTRYPOINT or CMD
statements supported by Docker. This way, Kliko is non-intrusive and can be easily added to existing containers that already
set an ENTRYPOINT or CMD.

\subsection{Expected runtime behavior}
During runtime, the Kliko runner will gather the parameters and expose them to the Kliko container. The content of the
variables is exposed by the Kliko runner in the \texttt{/parameter.json} file, which should contain a flat dictionary
in JSON syntax\footnote{\url{https://www.json.org}}. JSON and YAML are structurally very similar, but YAML is designed
to be more human-readable, hence our choice of YAML for the Kliko definition. Future versions of Kliko will support
both formats.

While reading this text, one might get confused by the context
of the file location (inside or outside the container). As a rule of thumb if a path in this text starts with a slash
(\texttt{/}) it is \textit{inside} the container.

If one or more of the parameters is a file, those will be exposed by the Kliko runner in the read-only
\texttt{/param\_files} folder during runtime. It is the responsibility
of the Kliko container to parse the \texttt{/parameters.json} file, perform potential the run-time housekeeping and
convert the parameter keys, values and/or files into an eventual command do be executed.

It is recommended to write logging to stdout and stderr. This makes it easier for the Kliko runner to visualize or
parse the output of a Kliko image.

\subsection{Flavors of Kliko Images}
\label{flavors}
We distinguish two flavors of Kliko containers, \textit{joined Input/Output} (read-write) and
\textit{split IO} (read-only). The style of container is specified in the \texttt{io} field in \texttt{/kliko.yml} file
inside the container, see \S \ref{klikoformat}.

The difference is the way the contained software interacts with the working data. In the case of \textit{split IO}
the Kliko runner exposes the input data to the container in the \texttt{/input} folder. This folder is read-only, to
prevent accidental manipulation of the data. The Kliko container is expected to write any output data into the
\texttt{/output} folder. The Kliko Runner will then handle this output data after the container reaches the end if its
lifetime. A \textit{split IO} Kliko container should always yield the same results for multiple independent runs
when presented with the same data and parameters (formally is called, ``having no side effects''). This is basically the
essence of the functional programming paradigm.

In the case of \textit{joint IO} there is only one point of interaction with the Kliko host, \texttt{/work} which is exposed
read/write. Basically, the input and output folders are combined into one that is mounted with read/write permissions.
Contrary to the \textit{split IO} flavor, this might be potentially dangerous for data processing as it can alter the
original data.

From a run-time parallelization perspective, the \textit{split IO} flavor is preferred. A container without side effects
enables the Kliko Runner to do graph-based logical inference of dependencies and execution scheduling, reuse results and
also run various containers in parallel, potentially resulting in faster execution. In practice, existing software
does not
always support this type of operation, or it is simply not feasible to create a copy of the data. In that case, the
\textit{joined IO} style has to be used.

\subsection{The \texttt{/kliko.yml} schema}
\label{klikoformat}

A \emph{kliko.yml} file is a YAML file and it \emph{should} contain the fields listed in Table~\ref{tab:klikofields}.

\begin{table*}[t]
\caption{Required Kliko fields}
\label{tab:klikofields}
\begin{tabular}{l | p{0.7\textwidth} }
\hline
field       & description \\ \hline
\textbf{schema\_version} & The version of the Kliko specification, independent of the versioning of the Kliko library \\
\textbf{name}    &   Name of the Kliko image. For example ``radioastro/simulator'' for RODRIGUES. \\
\textbf{description}  &  A more detailed description of the image. \\ 
\textbf{url}   & Website of project or repository where project is maintained  \\
\textbf{io}   & ``join'' or ``split''. See the two flavors of Kliko Containers in \S \ref{flavors} \\
\textbf{Sections} & a list of one or more sections, grouping fields together. 

\end{tabular}
\end{table*}

Each section  contains a list of fields. Each fields statement should contain a list of field elements. Each field
element has two mandatory keys, a name and a type. Name is a short reference to the field which needs to be unique.
This will be the name for internal reference. The type defines the type of the field, possible types are listed in
Table~\ref{tab:klikotypes}. Depending on the type there are optional extra fields, listed in Table~\ref{tab:klikotypefields}.

\begin{table*}[t]
\caption{Kliko variable types}
\label{tab:klikotypes}
\begin{tabular}{l | p{0.8\textwidth} }
\hline
type       & description \\ \hline
\textbf{choice} &   field with a predefined set of options, see the optional choices field below \\
\textbf{str}    &  string value \\
\textbf{float}  &   float value \\ 
\textbf{file}   &  A file path. This file will be exposed in \texttt{/param\_files} at runtime by the Kliko Runner  \\
\textbf{bool}   &  A boolean value \\
\textbf{int}    &  An integer value
\end{tabular}
\end{table*}

\begin{table*}[t]
\caption{Kliko field types}
\label{tab:klikotypefields}
\begin{tabular}{l | p{0.8\textwidth} }
\hline
field    & description \\ \hline
\textbf{initial} &   supply a initial (default) value for a field \\
\textbf{max\_length}    & define a maximum length in case of string type \\
\textbf{choices}  &  define a list of choices in case of a choice field. The choices should be a mapping \\ 
\textbf{label}   &  The label used for representing the field to the end user. If no label is given the name of the field is used  \\
\textbf{required}   &  Indicates if the field is required or optional \\
\textbf{help\_text} &  An optional help text that is presented to the end user next to the field \\
\end{tabular}
\end{table*}

The schema described above is defined in the Kwalify format. Kwalify is a parser and schema validator for YAML and
JSON\footnote{\url{http://www.kuwata-lab.com/kwalify}}.
The definition itself is also  written in YAML. The Kliko library pykwalify\footnote{\url{https://github.com/Grokzen/pykwalify}}
is used to validate the YAML file against a schema. The full Kliko version 2 schema is listed Listing~\ref{code:kliko.yml}
in \ref{appendix:example}.

\subsection{The \texttt{/parameters.json} file}

When a container is started, the Kliko runner will mount a \texttt{/parameters.json} file into the container. This file
contains all parameters for the container in the JSON format.  The \texttt{/kliko} script supplied by the container
author should read and parse the \texttt{/parameters.json file}. The Kliko library (\ref{sec:klikolibrary}) supports
helper functions and scripts to parse and validate this file.
Validation is done based on the \texttt{/kliko.yml} definition, which is useful for preventing or tracking down problems.

An example parameters file that could be generated based on the kliko.yml definition is shown in Listing
\ref{code:parameters.json}.

\begin{listing}[htbp!]
\begin{minted}[frame=single,fontsize=\scriptsize]{json}
{
   "int": 10,
   "file": "some-file",
   "char": "gijs",
   "float": 0.0,
   "choice": "first"
}
\end{minted}
\caption{Example \texttt{parameters.json} file}
\label{code:parameters.json}
\end{listing}

Note that the sections are not supplied since they are only used for grouping and representation to the user.

\section{Running Kliko containers}

\subsection{Running a container manually}

As an example, we will describe an extremely simple Kliko container named ``fitsimagerescaler", available on our github
repository described in \S \ref{sec:Software}. This container takes a FITS image file which resides in the
\texttt{/input} directory, opens it, multiplies the pixel values by a parameterized value (2 by default) and exports
the result as a new FITS image in \texttt{/output}. The actual code that is run in \texttt{/kliko} is shown in
Listing \ref{code:fitsimagerescaler}.

Starting the Kliko container is nothing more than starting the container using Docker with some specific flags. If the
\texttt{parameters.json} file already exists, starting the container from the command line looks like this:

\begin{listing}[htbp!]
\begin{minted}[frame=single,fontsize=\scriptsize]{console}
 $ docker run -t -i \
    -v `pwd`/parameters.json:/parameters.json:ro \
    -v `pwd`/input:/input:ro \
    -v `pwd`/output:/output:rw \
    kliko/fitsimagerescaler /kliko
\end{minted}

\caption{Command for running a Kliko container manually. The \texttt{/parameters.json} file is mounted as well as the
input and output directories, in the ``split'' mode. The \texttt{kliko.yml} is already inside the container.}
\label{code:manual_kliko}
\end{listing}

\texttt{`pwd`/input} and \texttt{`pwd`/input} are input and output folders in your current working directory
outside the container. \texttt{`pwd`} is required since the Docker engine can only work with absolute paths.

This command fires up the \texttt{fitsimagerescaler} container, mounts the \texttt{parameters.json} file as well as
input/output directories and runs the \texttt{/kliko} script located in the root directory. In our case, the FITS image
file has to be present in the local input directory for the script to run properly.

For the reader unfamiliar with Docker this command might look cumbersome and error-prone but the command constitutes the
fundamental principle of Kliko (in addition to specification and the extensive test suite). This can be used a base to create your
own Kliko runner in any language that has Docker bindings.

This way of implementing inputs, outputs and running generic scripts, demonstrates that it becomes fairly easy to
connect the input parameters and data (generated by scripting and/or a web form) to software living inside the
container. Kliko implements a set of tools to insure the robustness of this implementation.

\subsection{Inside the Kliko container}
\label{sec:klikolibrary}

The \texttt{/kliko} script is the first entry point into the specifics of the container. We can easily parse the
\texttt{/parameters.json} file using a JSON parser in python, by performing the commands in code listing
\ref{code:parse_params}.

\begin{listing}[htbp!]
\begin{minted}[frame=single,fontsize=\scriptsize]{python}
 import json
 parameters = json.load(open('/parameters.json', 'r'))
\end{minted}
\caption{Example of how to parse \texttt{parameters.json} file with standard python packages.}
\label{code:parse_params}
\end{listing}

However, at this point, the parameters file is not yet validated. We can be sure that the parameters file is actually
generated from our Kliko definition by installing the Kliko library inside the Kliko container and using it from our
\texttt{/kliko} script\ref{code:validate_params}. Validation helps reduce human or programming error.

\begin{listing}[htbp!]
\begin{minted}[frame=single,fontsize=\scriptsize]{python}
from kliko.validate import validate
parameters = validate()
\end{minted}
\caption{Example how to parse \texttt{parameters.json} using the Kliko library}
\label{code:validate_params}
\end{listing}

After the Kliko validation is performed, a dictionary is created and all values can be used freely inside the script
itself (by passing them to functions) or passed directly to the container OS as environment variables.
All this validation is intended to reduce human or programming error as early as possible.

\begin{listing}[htbp!]
\begin{minted}[frame=single,fontsize=\scriptsize]{python}
import kliko
from kliko.validate import validate
from astropy.io import fits

parameters = validate()
file = parameters['file'] # filename in /input
factor = parameters['factor'] # multiplying factor

print('welcome to fits multiply!')
print("'%s' multiplied by '%s':" % (file, factor))

data = fits.getdata(file)
multiplied = data * factor
output = path.join(kliko.output_path, path.basename(file))
fits.writeto(output, multiplied, clobber=True)
\end{minted}
\caption{Example of \texttt{/kliko} file scale the values in a FITS image}
\label{code:fitsimagerescaler}
\end{listing}

\subsection{kliko-run}
Instead of calling Docker directly, Kliko is bundled with \texttt{kliko-run}, a command line utility that enables a user
to run a Kliko container in a seamless way. It also assists in exploring the parameters that a given Kliko container
supports. Code listing \ref{code:kliko-run} shows the docstring of the kliko-run command for a simple container
(available as a test container shipped with Kliko). The optional arguments are generated automatically from the YAML
file. It shows how any shipped application can be easily interfaced with the host system, in a way that part (or all)
of the variable names of the application can be modified directly from the command line.
This enables Kliko, with the help of Docker, to ship a complex software as an application that is equipped with a
simple interface. Kliko provides a simple way to implement this interface in a controlled and robust way while being
completely agnostic about the mechanics happening inside the container.

\begin{listing}[htbp!]
\begin{minted}[frame=single,fontsize=\scriptsize,breaklines]{console}
$ kliko-run kliko/fitsimagerescaler --help

   usage: kliko-run [-h] [--target_folder TARGET_FOLDER] --choice {second,first}
                    --char CHAR [--float FLOAT] --file FILE --int INT
                    image_name

   positional arguments:
     image_name

   optional arguments:
     -h, --help            show this help message and exit
     --target_folder TARGET_FOLDER
                           specify output or work folder (default: ./output)
     --choice {second,first}
                           choice field (default: second)
     --char CHAR           char field, maximum of 10 chars (default: empty)
     --float FLOAT         float field (default: 0.0)
     --file FILE           file field, this file will be put in /input in case
                           of split io, /work in case of join io
     --int INT             int field
\end{minted}
\caption{Output of the \texttt{kliko-run} command}
\label{code:kliko-run}
\end{listing}

\section{Chaining containers}

Kliko containers can also be chained. Chaining means that the output of a container is connected to the input
of a consequetively executed container. This enables the creation of workflows. Additionally, if the Kliko containers
are ``split IO", we can execute containers that do not depend on each other in parallel. Their intermediate results
can be cached, which can speed up execution time of future workflow runs and can help debugging problems with the
workflow by examining intermediate results.

There are various workflow creation frameworks and libraries available. We evaluated two popular Python-based workflow
management libraries, airflow\footnote{\url{https://airflow.apache.org}} and
Luigi\footnote{\url{https://github.com/spotify/luigi}}. Although Kliko is designed to be workflow management independent, Luigi is a
better fit. Airflow is intended to visualize automated repetitive tasks like cron jobs, while Luigi is more
oriented towards once-off batch processing. Luigi is an open source Python library that handles dependency resolution,
does workflow management, optionally visualizes data in a web interface and can handle and retry failures. At the
core of a Luigi workflow is the Task, which is a Python class that defines what to be executed, how to check if this
task has already been executed and optionally if it depends on the result of another task. This is a very simple
but powerful concept that integrates fluently with Kliko. The Kliko library contains a KlikoTask definition which
can be used to integrate Kliko in a Luigi pipeline.

\section{Example usage of Kliko}
\subsection{VerMeerKAT}

\begin{figure}[!htbp]
\centering
\includegraphics[width=3in]{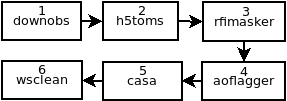}
\caption{Flow diagram of the VerMeerKAT data reduction pipeline.}
\label{fig:vermeerkat}
\end{figure}

To illustrate the mechanics of chaining container together we explain a real world application here, the VerMeerKAT
pipeline.

VerMeerKAT is a semi-automated data reduction pipeline for the first phase of deployment of the MeerKAT
telescope\footnote{http://www.ska.ac.za/gallery/meerkat/}\cite{booth2009meerkat}\cite{jonas2009meerkat}. All steps in this
pipeline are shown in
Figure \ref{fig:vermeerkat}. It is a closed source project used internally at SKA South Africa and is based on a set of
bash scripts. Using bash for this is not ideal; it is hard to make a portable pipeline, not trivial to recover and
continue from errors, reuse intermediate results. Parallelization is possible, but this needs to be manually and
explicitly defined in the scripts.

For this paper we made a Kliko version of this pipeline\footnote{https://github.com/gijzelaerr/vermeerkat-kliko}.
Using Kliko for composing this pipelines has some key advantages; i) easy installation and deployment of the software
ii) optional caching of  intermediate data products; ii) implicit parallelization of tasks independent steps; and iii)
progress visualization and reporting using a tool like Luigi.

The VerMeerKAT pipeline, (see Figure \ref{fig:vermeerkat}), starts by querying the MeerKAT data archive for a given set of
observations (step 1), along with the meta-data for that observation. Next step is to convert the downloaded data from the hdf5
format to a MeasurementSet (MS)\cite{kemball2000measurementset} (step 2), since most radio astronomy tools only support this
format. Once the data is in the MS format, it is then taken through a series of
manual and automated tools that excise data points that are contaminated by radio frequency interference
\cite{Prasad2012,aoflagger} (step 3, 4 and 5). The data are then calibrated \cite{hamaker2006,oms2011b} and imaged \cite{white7} (step 6).

For the creation of the VerMeerKAT Kliko containers, we make use of the packages from KERN\cite{KERN}. KERN is a
bi-annually released set of radio astronomical software packages. This suite contains most of the tools that a
radio astronomer needs to process radio telescope data. These packages are precompiled binaries in the Debian format
and contain all the metadata required for installing the package like dependencies and conflicts. KERN is only
supported on Ubuntu 16.04 at the moment of writing, but that is no problem inside a Docker container.

Listing \ref{code:kern} is an example Dockerfile for the wsclean\cite{offringa2014wsclean} Kliko container in VerMeerKAT.
When this file is built it will install the KERN package of wsclean inside the container and bundle the container with
the Kliko definition and parser script. The Docker files for the other steps are very similar.

\begin{listing}[ht]
\begin{minted}[frame=single,fontsize=\scriptsize,breaklines]{console}
FROM kernsuite/base:1
RUN docker-apt-install wsclean
ADD kliko.yml /
ADD kliko /
\end{minted}
\caption{Dockerfile for a KERN package}
\label{code:kern}
\end{listing}

\begin{figure}[!htbp]
\centering
\includegraphics[width=3in]{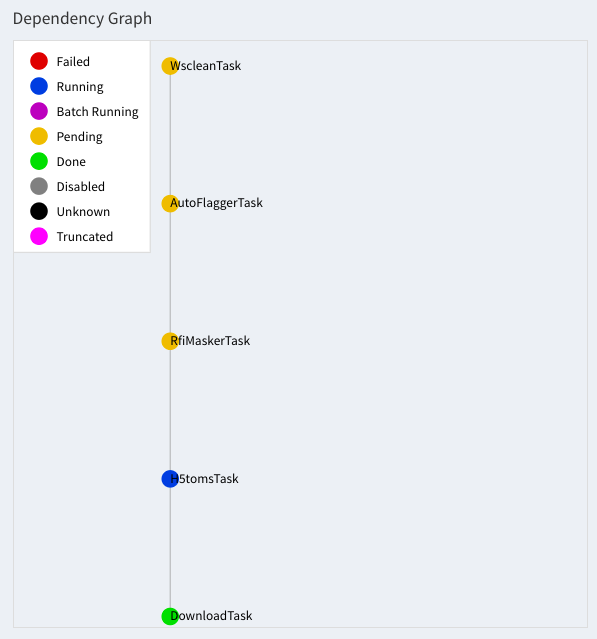}
\caption{Screenshot of Luigi, running the vermeerkat pipeline }
\label{fig:luigi}
\end{figure}

Listing \ref{code:klikotask} is an example Kliko task definition. This example will use the rfimasker Kliko
containers. It depends on the H5tomsTask Kliko task. When this task is invoked using Luigi, Luigi will do the dependency
resolution, check if the required tasks have run and if not, run them. The progress can be visualized with the Luigi
interface (Figure \ref{fig:luigi}). All other steps in the workflow are very similar to this example.

\begin{listing}[ht]
\begin{minted}[frame=single,fontsize=\scriptsize,breaklines]{python}
from kliko.luigi_util import KlikoTask

class RfiMaskerTask(KlikoTask):
    @classmethod
    def image_name(cls):
        return "vermeerkat/rfimasker:0.1"

    def requires(self):
        return H5tomsTask()
\end{minted}
\caption{An example \texttt{KlikoTask}}
\label{code:klikotask}
\end{listing}

\subsection{RODRIGUES}
\label{rodrigues}
\begin{figure}[!htbp]
\includegraphics[width=0.9\columnwidth]{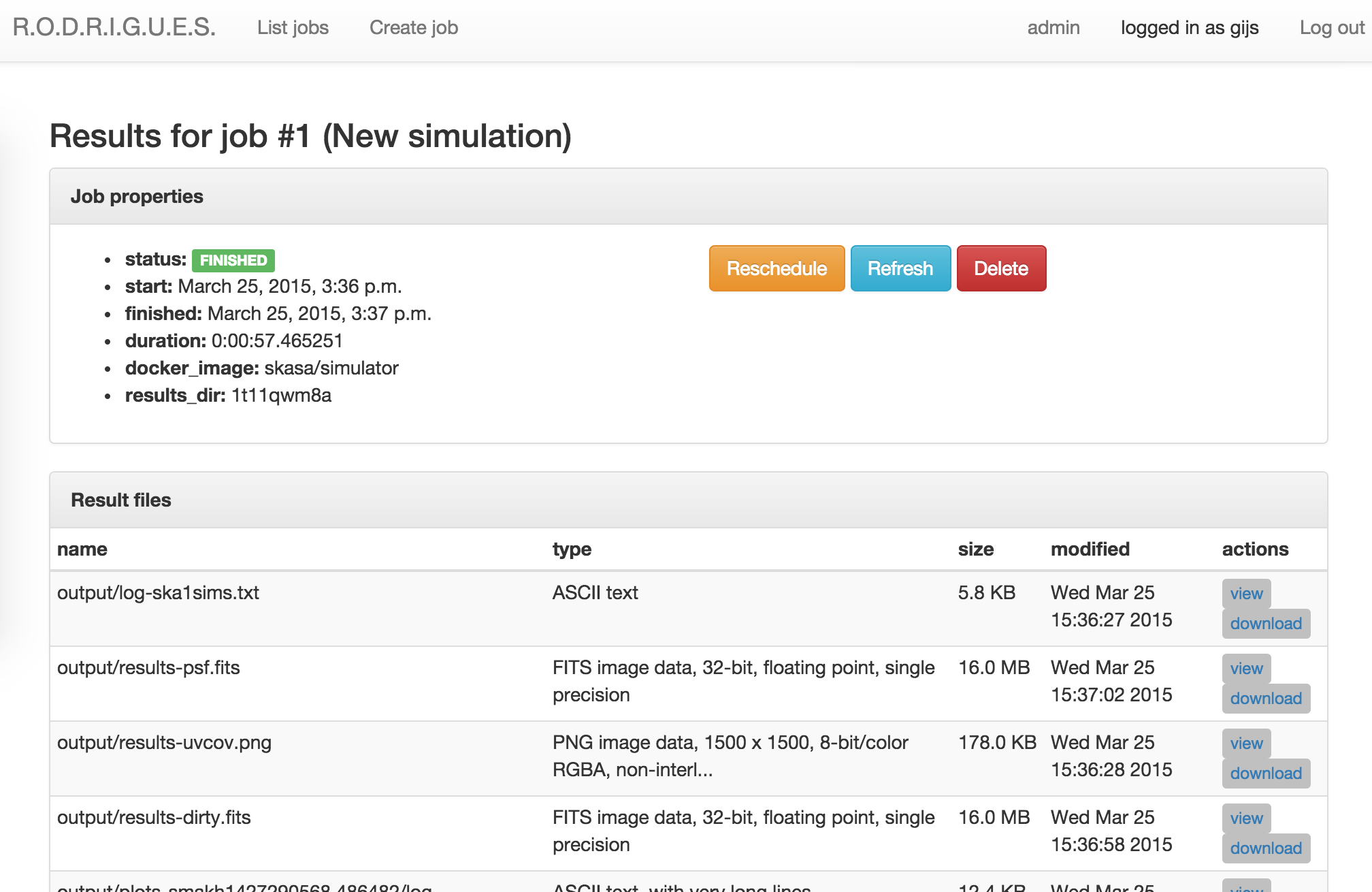}
\caption{Screenshot of RODRIGUES, the result visualizer.}
\label{fig:rodrigues_result}
\end{figure}

\begin{figure}[!htbp]
\includegraphics[width=0.9\columnwidth]{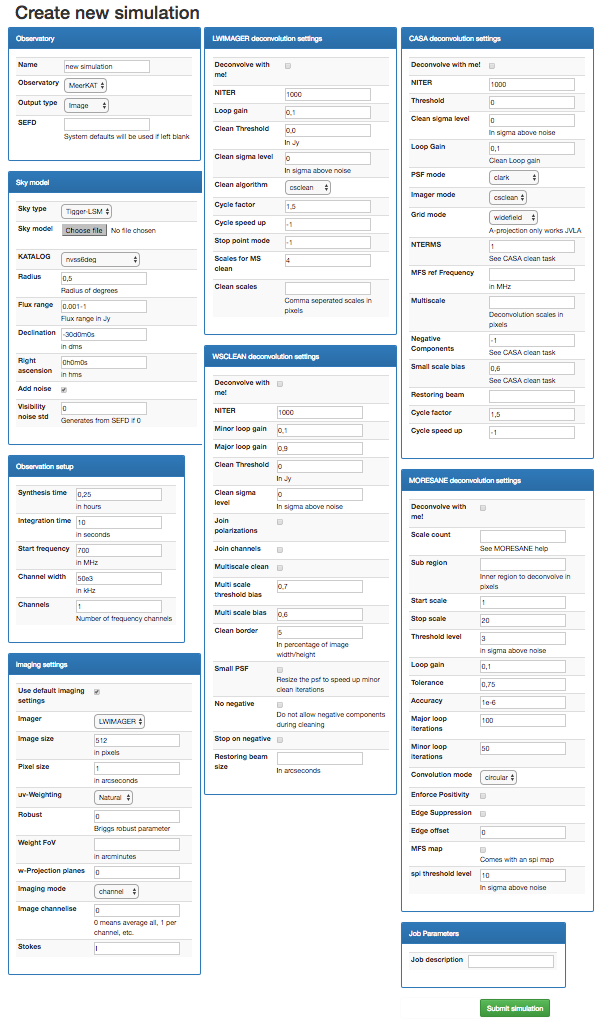}
\caption{Screenshot of RODRIGUES, parameter form is generated from Kliko definition.}
\label{fig:rodrigues_form}
\end{figure}

Another project using Kliko is RODRIGUES (RATT Online Deconvolved Radio Image Generation Using Esoteric
Software)\footnote{\url{https://github.com/ska-sa/rodrigues}}.
RODRIGUES is a web-based Kliko job scheduling tool and it uses the Kliko as a required format for the job.
RODRIGUES acts as a ``kliko runner''. A user of RODRIGUES can log into RODRIGUES and add a new
Kliko container. RODRIGUES will open the Kliko container, parse the parameters and expose these parameters to
the user using a web form (Figure~\ref{fig:rodrigues_form}). The user can then fill in the parameters in this form
and submit the job into the RODRIGUES container queue. The container will be run on the system configured by the
RODRIGUES system administrator. Once the job is finished the results are presented to the user in the same web interface
(Figure~\ref{fig:rodrigues_result}).

RODRIGUES makes it much easier to schedule new jobs with varying parameters, enabling scientists with minimal
programming or computing knowledge to run experiments in a clean, visual, structured and reproducible way.

\section{Software availability}
\label{sec:Software}
Kliko and its library are open source, licensed under the GNU Public License
2.0\footnote{\url{https://github.com/gijzelaerr/kliko}}. Kliko is bundled with an extensive test suite which covers
80\% of the source code as of the current release \texttt{0.6.1}. The Kliko library is written in Python and is
compatible with Python 2.7, all Python 3 versions and even PyPy. Development and distribution is done on Github and a
third party continuous integration service runs the full test suite on all supported platforms for every commit and every
Github pull request.

\section{Discussions and Prospects}
\subsection{Limitations}
\label{sec:Limitations}
While developing Kliko, we ran into various issues with Docker which might limit the applicability. It is up to the user
to decide if this affects the usefulness of Docker and or Kliko. These issues are listed hereafter for your consideration.
That said, the field of containerization is evolving very fast and hopefully most of these issues will be resolved soon
or can be worked around.

First of all, being able to run a Docker container on a system is very similar to giving the user administrative
access to the machine, that is the user can escalate easily to root privileges
\footnote{\url{https://github.com/docker/docker/issues/6324}} \cite{Bui2015Analysis}. The singularity containerization
technology has a more secure design and we are planning to add support for this framework in future versions of
Kliko.

Additionally, using GPU acceleration with NVidia hardware is not trivial, since the kernel driver version and library
version need to match up, breaking the independence between host and container. There is a workaround available, but
this requires a replacement of the Docker daemon with a custom one \footnote{\url{https://github.com/NVIDIA/nvidia-docker}}.

A similar issue arises with optimization flags. For example, SIMD instructions can greatly enhance the run-time speed,
but not all x86 processors support all SIMD optimization. This will result in crashes of the binary if it is compiled
with optimization not supported by the host. Again this breaks the platform independence assumption. A good strategy
is to be conservative and compile your binaries for the oldest architecture you intend to support. The good news is
that it is easier to support multiple target platforms in the same binary when using modern versions of GCC \footnote{\url{https://lwn.net/Articles/691932}}.

Another issue is that it is easy to inherit Docker definitions from other Docker definitions, but it is currently not
possible to combine Docker definitions together or to inherit from multiple Docker definitions at the same time (merge).
Following the Docker philosophy, Docker containers should have a single responsibility, so this should not be a problem,
it does not require mixing of Docker definitions. However, in practice,
this is not always possible: sometimes various big libraries need to communicate in the same memory space so a new
Docker container with all software needs to be created. The results are far away from minimal small single purpose
Docker containers.

For network intensive applications Docker may be less well suited, since the use of network address
translation\cite{Felter2015Updated}. Also here there is a workaround available by disabling the translation and using
the host network stack directly.

\subsection{Future Work}

During the development and usage of Kliko we became aware of CommonWL\cite{Amstutz2016}, a more generic approach to
describe applications input/output flow and parameters. We will investigate how we can incorporate CommonWL into Kliko
to extend the usability and user base.

At the moment Kliko is designed with other container solutions in mind.
Singularity is an alternative that looks to be gaining momentum within the scientific compute (HPC) 
community since it is more aware and careful of the security implications that come by allowing running
containers on a shared infrastructure.

\subsubsection{Streaming Kliko}

Kliko was born in the field of radio astronomy. Most tools in this field operate on data living on disk.
Radio astronomy uses several file formats, for example, casacore Measurement Sets, FITS images and, in
some cases HDF5. Using the file system is an easy to understand and technically stable approach,
but problems arise when the size of the dataset grows. Emerging telescope arrays such as MeerKAT, 
followed by SKA phase 1, will result in an exponential growth in data rates. Repeatedly reading and writing 
data from to slowest medium in a computer -- the disk -- is not going to scale and a new strategy is needed.
Directly streaming data between processing tasks will be required. Although this is already being done \footnote{\url{https://github.com/ska-sa/spead2}} in some
pipelines, there is no field-wide accepted standard that fits all needs. Our plan of action is to investigate
existing solutions being used, investigate industry standards \footnote{\url{https://github.com/google/protobuf}},
optionally create a sub-specification and open source reference implementation with support libraries for the most
used public languages\footnote{\url{http://arxiv.org/pdf/1507.03989.pdf}}.

\section{Conclusions}

Kliko is a Docker-based container specification. It is used to create abstract descriptions of the input
and output of existing software resulting in Kliko containers. These Kliko containers can be used to encapsulate a single
job or can be chained together in a pipeline. In the future, we probably will adopt the CWL standard to extend the
interoperability with other existing workflow tools. Kliko is written in Python, open source and available free to use.

\section{Acknowledgments}
J. Girard acknowledges the financial support from the UnivEarthS Labex program of 
Sorbonne Paris Cit\'e (ANR-10-LABX-0023 and ANR-11-IDEX-0005-02). The research of
O. Smirnov is supported by the South African Research Chairs Initiative of the Department of Science and 
Technology and National Research Foundation.

\section{References}

\bibliography{kliko}{}
\bibliographystyle{plain}

\newpage

\appendix
\section{The Kliko validation specification}
\label{appendix:spec}

\begin{listing}[!h]
\begin{minted}[frame=single,fontsize=\scriptsize]{yaml}
schema;fields:
 type: map
 mapping:
   name:
     type: str
     required: True
   type:
     type: str
     required: True
     enum: >
       ['choice', 'char', 'float', 'file', 'bool', 'int']
   initial:
     type: any
     required: False
   max_length:
     type: int
     required: False
   choices:
     type: map
     required: False
     mapping:
       regex;(.*):
         type: str
   label:
     type: str
     required: False
   help_text:
     type: str
     required: False
   required:
     type: bool
     required: False
type: map
mapping:
 schema_version:
   type: int
 author:
   type: str
 name:
   type: str
 description:
   type: str
 container:
   type: str
   pattern: .+/.+
 email:
   type: str
   pattern: .+@.+
 url:
   type: str
   pattern: >
     https?:\/\/(www\.)?[-a-zA-Z0-9@:
     %._\+~#=]{2,256}\.[a-z]{2,6}\b([-a-zA-Z0-9@:%_\+.~#?&//=]*)
 io:
   type: str
   required: True
   enum: ['split', 'join']
 sections:
   type: seq
   matching: "any"
   sequence:
     - type: map
       mapping:
         name:
           type: str
           required: True
         description:
           type: str
           required: True
         fields:
           type: seq
           required: True
           sequence:
             - include: fields
\end{minted}       
\caption{The Kliko definition version 2}
\label{code:klikodef}
\end{listing}

\newpage

\section{An example kliko.yml file}
\label{appendix:example}

\begin{listing}[!h]
\begin{minted}[frame=single,fontsize=\scriptsize]{yaml}
schema_version: 2
name: Kliko test image
description: for testing purposes only
container: kliko/klikotest
author: Gijs Molenaar
email: gijsmolenaar@gmail.com
url: http://github.com/gijzelaerr/kliko
io: split

sections:
 -
   name: section1
   description: The first section
   fields:
     -
       name: choice
       label: choice field
       type: choice
       initial: second
       required: True
       choices:
         first: option 1
         second: option 2
     -
       name: char
       label: char field
       help_text: maximum of 10 chars
       type: char
       max_length: 10
       initial: empty
       required: True
     -
       name: float
       label: float field
       type: float
       initial: 0.0
       required: False
 -
    name: section2
    description: The final section
    fields:
      -
        name: file
        label: file field
        help_text: a helpful text
        type: file
        required: True
      -
        name: int
        label: int field
        type: int
        required: True
\end{minted}       
\caption{Example \texttt{/kliko.yml}}
\label{code:kliko.yml}
\end{listing}

\end{document}